# ANALYSIS OF EXPERIMENTAL CONDITIONS FOR SIMULTANEOUS MEASUREMENTS OF TRANSPORT AND MAGNETOTRANSPORT COEFFICIENTS OF HIGH TEMPERATURE SUPERCONDUCTORS

**J. Mucha[1], S. Dorbolo[3], H. Bougrine[3], K. Durczewski[1],**

**M. Ausloos[2]**

[1]Institute of Low Temperature and Structural Research, Polish Academy of Sciences, PL-50-950 Wroclaw, Poland,

[2]SUPRAS, Institut de Physique B5, Universite de Liege, B-4000 Liege, Belgium

[3] SUPRAS, Institut d Electricite Montefiore B 28, Universite de Liege, B-4000 Liege, Belgium

*Corresponding author*:

Dr Jan Mucha
Institute of Low Temperature
and Structure Research  Polish
Academy of Sciences.
P.O. Box 1410  50-950 Wroclaw  Ok lna 2
tel: +48 71 34 350 25, fax: +48 71 344 10 29
 mailto:ja_mu@int.pan.wroc.pl28

**ABSTRACT**

Experimental conditions for simultaneous measurements of transport coefficients of high temperature superconductors in zero and non-zero magnetic fields are analysed. Test measurements of the thermal conductivity, the thermoelectric power and the Nernst - Ettingshausen effect of a textured Bi2212 sample are reported in an external magnetic field of 2T. Errors related to parameters of the thermocouple used and to the spurious heat flows are discussed for a new experimental set-up built based on a closed cycle helium refrigerator. Possible optimising of experimental conditions is suggested.

**INTRODUCTION**

The thermal conductivity, $k$, the thermoelectric power or Seebeck coefficient, $S$, and the Nernst-Ettingshausen coefficient, $N$, are the transport coefficients (and the magnetotransport coefficient in case of $N$) which play an important role in the discussion of the phenomenon of high temperature superconductivity. In the normal phase ($T > T_c$) they provide information on the electron scattering. In the superconducting phase ($T < T_c$), $S$ vanishes more or less rapidly. The thermal conductivity $k$, and the Nernst-Ettingshausen coefficient $N$, remain of a non-zero magnitude. In the case of $k$ this is a manifestation of the transport of heat by phonons and normal electrons, while in the case of $N$ it results from a manifestation of the movement of the vortices under the influence of the temperature gradient. The simultaneous measurement of these three coefficients in the normal and superconducting phase guarantees that they are measured under the same thermodynamical conditions and can be rightfully compared. This, in turn, permits to provide a more reliable interpretation, since, e.g., systematic errors are minimised and the behaviour of one coefficient may suggest what physical mechanisms are responsible for the behaviour of other ones. Furthermore the measurements can be put in parallel on the simultaneous ones of $k$, $S$ and the thermal diffusion, as described in [1-4].

The present paper is devoted to the implementation of an experimental procedure for the simultaneous measurement of $k$, $S$ and $N$. We shall describe the principles of the measurements, the experimental set-up and the measurement procedure. A textured sample of $Bi_2Sr_2CaCu_2O_8$ is used for testing.

The sample used for the test measurements was magnetically textured. The synthesis of the samples is described elsewhere [5]. Their texturation has been confirmed by X-ray diffraction. The separately measured transport coefficients of such samples textured in the same way have been already studied in [6]. Their behaviour manifested the mutual orientation of copper oxide planes resembling that of monocrystaline samples.

## EXPERIMENTAL PROCEDURE

A scheme illustrating the measurement method is shown in Fig. 1. The stream of heat passing along the sample is perpendicular to the surface area $A$. The temperatures $T_1$ and $T_2$ are measured at the marked points (see also Fig.2) . These temperatures and the distance $\Delta x$ between the points allow us to calculate the mean temperature of the sample˚: $T_s = (T_1 + T_2)/2$ and the thermal gradient on the sample˚: $\Delta T/\Delta x$. The electric field gradients need to be measured in directions perpendicular to the magnetic field : $E_x$ along the temperature gradient and the heat stream, and $E_y$ perpendicular also to the latter direction (see fig.2) in order to obtain $N$ and the field dependence of $k$ and $S$. If the power $Q$ of the heater producing the heat stream in the x-direction and the strength of the external magnetic field in z-direction are known , one has˚:

$$k\ (T_s) = (\Delta x\ Q)\ /\ (A\ \Delta T\ ) \tag{1}$$

$$S\ (T_s) = (E_x\ \Delta x)\ /\ \Delta T \tag{2}$$

$$N\ (T_s) = (E_y\ \Delta x)\ /(\ B_z\ \Delta T) \tag{3}$$

Figure 1 shows schematically the configuration of electrical, thermal and magnetic fields, which realises the above idea. The system is built on the cold head of a Gifford-McMahon (GM) closed cycle helium refrigerator being used as a cryostat. The connection of the sample to the GM refrigerator as well as the connection of the small metallic heater (150 Ohm) to the sample are made with GE 7031 varnish. The sample was enclosed within a radiation shield of Al foil plated with Cu thermally anchored to the cold head. The dynamical vacuum in the measurement chamber was as low as $10^{-6}$ to $10^{-7}$ mmHg. Therefore, the influence of the residual gases and of gas desorption on results of our measurements can be excluded In order to improve the efficiency of the sample shielding a heater is wound around the shield. The power of this heater is electronically controlled so that the sample-heater differential thermocouple indicates a value close to zero. We tested various configurations and found that the shielding efficiency is optimal when the temperature dependence and the magnitude of the thermal conductivity of the material of the shield and the sample are nearly the same. This is particularly important for thermal measurements above 100 K . The temperature of the shield was electronically adjusted to that of the sample with the accuracy 0.01 K. More detailed information on the conditions of sample shielding in measurements of thermal properties of materials at low temperatures can be found in [7,8]. The temperature of the cold head, $T_0$, is stabilised by a Lake Shore Temperature Controller DRC - 91 in the range 30K to 300K with a $5\times10^{-3}$ K accuracy on the hour scale. All wires used are anchored in such a way that the parasitic heat flow is smaller than 0.1 % of the power produced by the sample heater.

Where to anchor the sample heater is also important The chromel, Cr, constantan, Ct, and Cu wires of the 50μm caliper are point — welded and soldered to the sample. Au wires are used to measure the Nernst voltage $V_N$, while Cu wires are used to measure the Seebeck coefficient. Two phosphor-bronze wires of the same caliper are lead to the sample heater to measure the voltage ($V^+,V^-$) and two copper wires of the 70 μm caliper are used to supply current ($i^+,i^-$) to the heater from a stabilised current source 224 Keithley. When the temperature gradient along the sample exceeds 1 K, it is better to anchor a shield and the sample on the cold head (see fig.1) - in order to eliminate the parasitic heat flow, since the temperature of the shield should be the same as that of the sample [8] . During measurements with small gradients the heater wires need to be anchored to the electronically stabilised cold head.

**EXPERIMENTAL RESULTS**

The new built experimental system was successfully used for precise measurements of the thermal conductivity, the thermoelectric power and the Nernst - Ettingshausen effect. Typical collected results on a magnetically textured Bi-2212 sample are shown in Figures 3 and 4 for zero and 2 T magnetic field, respectively.

The measured thermal conductivity, $k$, is a smooth function of temperature(Fig. 3). The room temperature value of $k$ is about 6W/Km and corresponds to that in other reports on Bi-2212 single crystal superconductors [9]. At zero magnetic field $k$ exhibits a broad plateau above 100 K. Below the critical temperature $T_c = 87$ K the thermal conductivity increases abruptly with decreasing temperature. Such a temperature variation of thermal conductivity is characteristic of high quality superconductors. A very similar behaviour was also observed in the *ab* plane of the textured Y123/Y211 superconductors [10]. When measured in a 2 T magnetic field the temperature variation of $k$ in the normal state becomes more flattened. The *k* minimum is located near T$_c$ and *k* increases up to about 11.0 W/Km at 30 K again (Fig. 4) when the temperature is lowered.

The normal state thermoelectric power diminishes slowly from about 16 µV/K at 220 K at a mean rate of 0.05µV/K$^2$ for a temperature decrease. The superconducting transition occurs in the 76 -89 K interval (Fig. 3). When the 2 T magnetic field is applied the normal state thermoelectric power is not altered but the transition interval broadens up to about 53 - 100 K (Fig. 4).

It is known that the Nernst - Ettingshausen effect creates a very weak voltage signal both in the normal and mixed states of high temperature superconductors. Our results presented in Fig.4 confirm this standard behaviour [1]. The normal state Nernst - Ettinghausen coefficient is a slowly decreasing function of temperature with a mean slope of 0.0055 µV/TK$^2$ in agreement with other observations [11]. In the mixed state the Nernst-Ettingshausen coefficient behaves also in the standard way. It rises up to the maximum with increase of the temperature, then lowers and merges with the values in the normal state. Since our textured Bi-based sample was used only for testing our set-up we confine ourselves to the above description and refer the reader for details and the interpretation to results for similar material , which we described in [12].

**ESTIMATES OF EXPERIMENTAL ERRORS**

The present analysis of the experimental errors is made in the same way as in [13], where problems arising from the application of thermocouples at low temperatures are described in detail. The relative error of the temperature measurement can be estimated as˚:

$$\varepsilon = \Delta T(0) / (T_s - T_0) \cong F(l/\alpha) \tag{4}$$

where $\Delta T(0)$ is the temperature jump at the contact of the thermocouple and the sample, $T_s - T_0$ is the temperature difference between the sample and the heat reservoir. $F(l/\alpha)$ is a function of two arguments: $l$ - the length of the contact of the thermocouple with the sample and $\alpha$ the characteristic length of the thermocouple connection. If $l >> \alpha$ this function can be represented as [12]:

$$F(l/\alpha) \approx 2(\alpha/L)exp(-l/\alpha) \tag{5}$$

where $L$ is the length of the thermocouple wire measured from the sample to the heat reservoir. The characteristic length of the thermocouple is determined as:

$$\alpha = ( k_w A_w t / k_B d )^{1/2} \qquad (6)$$

where $k_w$ is the thermal conductivity of the thermocouple wire of the cross section area $A_w$, $k_B$ is the thermal conductivity of the cement bond joining the thermocouple with the sample. The $t$ and $d$ are the thickness and width of the cement bond, respectively. Notice that $\alpha$ is a function of the temperature and the thermocouple — sample contact quality. A higher quality contact is indicated by smaller values of $\alpha$. In our measurements we used the Cr - Ct thermocouple with the caliper 0.05 mm and indium as bond. Since the thermal conductivity of Cr and Ct can be assumed to be nearly the same, the characteristic length of $\alpha$ is about 0.8 mm at 300 K. We mention for comparison that the value of $\alpha$ is as large as several mm at room temperature [12]. Thus, the condition $l >> \alpha$ is well satisfied in the case of our set-up. In order to determine the error of the temperature measurement we can use equations (5) and (6). For $\alpha = 0.8$ mm, $l = 5$ mm, $L = 100$ mm we obtain $\varepsilon \cong 0.003\%$. This means that such an error is negligible.

If indium is replaced by some other bonding material with reduced thermal conductivity, the values of $\alpha$ increase. In the extreme case of GE7031 varnish $\alpha$ may be several times higher. This in turn causes that an error in temperature determination may reach about 1.5% in our experimental set-up. This inaccuracy in temperature determination involves further errors of the same order of magnitude in the calculation of the thermoelectric power, Nernst - Ettingshausen coefficient and thermal conductivity.

A second source of errors is due the spurious heat flow, which may affect values of the thermal conductivity. The set of wires is responsible for the spurious heat flows since the heat stream $Q$ produced by the sample heater (see Figs.1,2 ) does not pass only along the sample but partially escapes along these wires. The thermal resistivity of the contacts $W_c \approx \alpha /( A_w k_w )$ and that of the wires $W_w = L / (A_w k_w)$ determine the amount of escaping heat. Since $W_w >> W_c$, the value of $W_w$ should be in fact taken into account to estimate the heat flowing along the set of wires . In our measurements the magnitude of $W_w$ is large and the relative error due to the heat flow along the set of wires is of the order 0.1% of $Q$. The geometrical error (of cross-section area of the sample - $A$ and the distance - $\Delta x$ of the anchoring of the thermocouple) does not exceed 4%. This allows us to estimate the total random error to be less than – 1% in the case of the thermal conductivity measurement without magnetic field.

When measurements of the thermoelectric and Nernst — Ettingshausen are made in a magnetic field, variable spurious voltages are generated in the thermocouples and wires due to mechanical vibrations originating from the Gifford-McMahon system. The total random error increases up to – 2.5 % when a 2 T magnetic field is applied. This magnetically generated contribution may be suppressed by minimising the wire lengths and locating the wires parallel to the direction of the magnetic field. On the other hand, shorter wires facilitate the spurious heat flows affecting the sample. This, in turn, additionally perturbs the thermal conductivity values for very short wires.

**CONCLUSIONS**

Simultaneous measurements of the transport properties of high temperature superconductors are known in the literature [3,7]. They are performed to accelerate the data collection and comparison. However, up to now no simultaneous measurements of $k(T)$, $S(T)$ and $N(T)$ have been performed. Our method of the simultaneous measurements of these three coefficients has two advantages: (i) it allows to obtain the final results in shorter time and (ii) to measure the three coefficients in the same temperature gradient at every temperature applied to the sample. Notice finally that the assembling and disassembling of the sample changes the density of the physical defects in the sample. It is known [14] that the density of the defects influence the transport coefficients and mostly the thermal conductivity. The experimental data collected during the simultaneous measurements of the thermal conductivity, the thermoelectric power and the Nernst - Ettingshausen effect of a textured Bi2212 superconductor confirm that an error arising from thermocouple connectors may be negligibly small. The analysis shows that the heat streams along the wires contribute much to the errors on the calculation of the thermal conductivity. The errors increase when measuring in a magnetic fields due to spurious voltages originating from vibrations of the Gifford — McMahon refrigerator. Experimental conditions may be optimised by a proper choice of wire lengths and location.

**Acknowledgements.** Authors thank Professor H.W. Vanderschueren for allowing to use the M.I.E.L. equipment and his constant interest. Authors are grateful to Dr. M. Pekala for reading the manuscript and critical remarks. Work was partially supported through PST.CLG.977377 NATO grant and Belgium - Poland Scientific Exchange Agreement.

**FIGURE CAPTIONS**

Fig.1. Schematic illustration showing the directions of electrical, thermal and magnetic fields used to determine the thermal conductivity, the thermoelectric power and the Nernst - Ettingshausen effect. Note that the heat flow is opposite to the temperature gradient.

Fig.2. Experimental set-up for the simultaneous measurement of the thermal conductivity,the thermoelectric power and the Nernst - Ettingshausen effect at low temperature.

Fig.3. Thermal conductivity and thermoelectric power versus temperature for Bi2212 superconductor without magnetic fields.

Fig. 4. Thermal conductivity, thermoelectric power and Nernst - Ettingshausen coefficient versus temperature for Bi2212 superconductor at 2T.

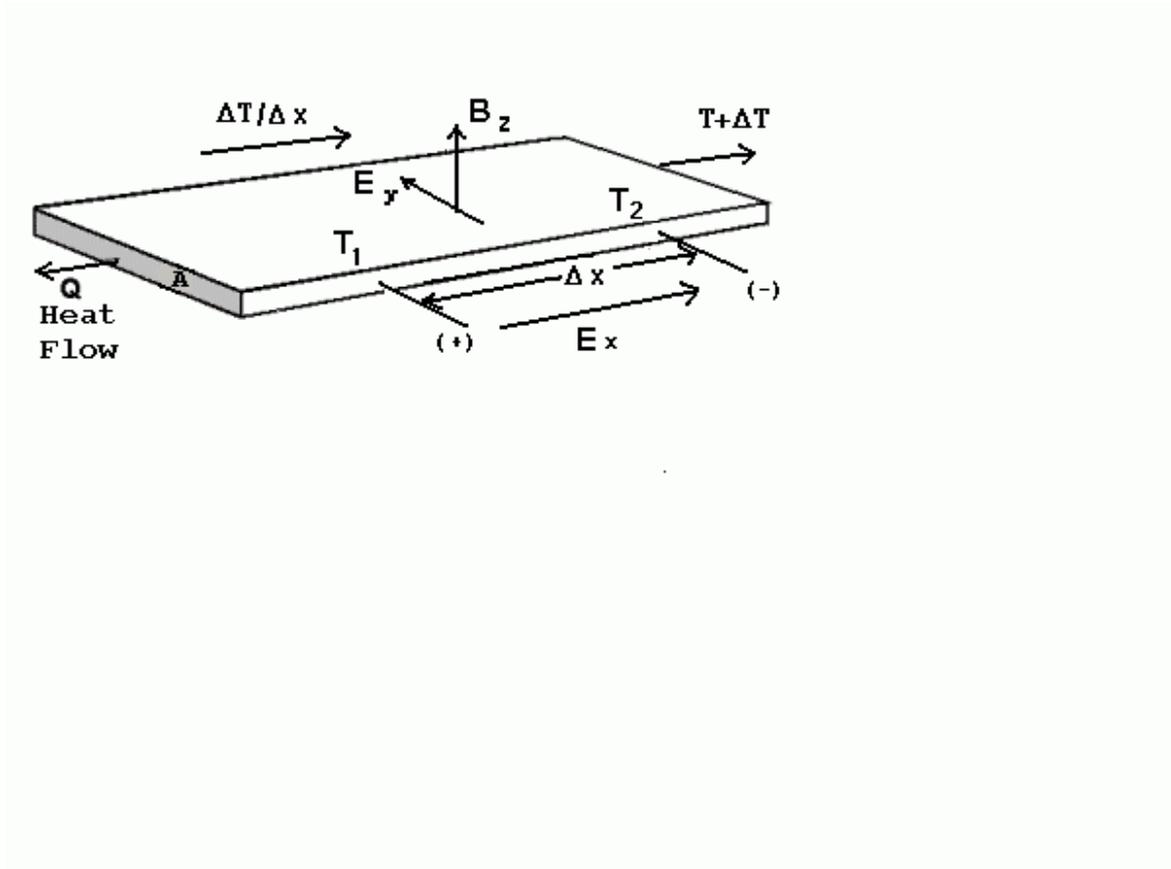

Fig.1

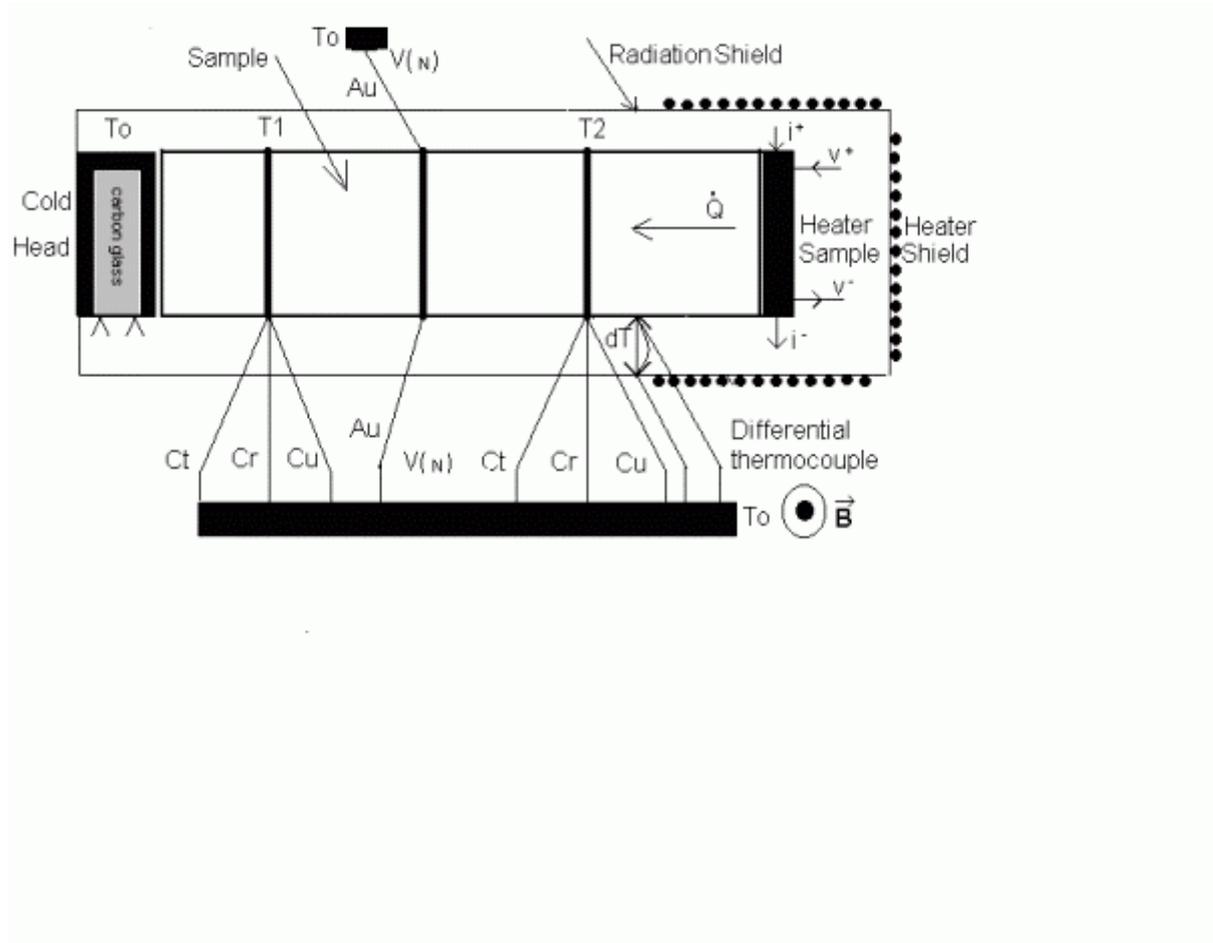

Fig.2

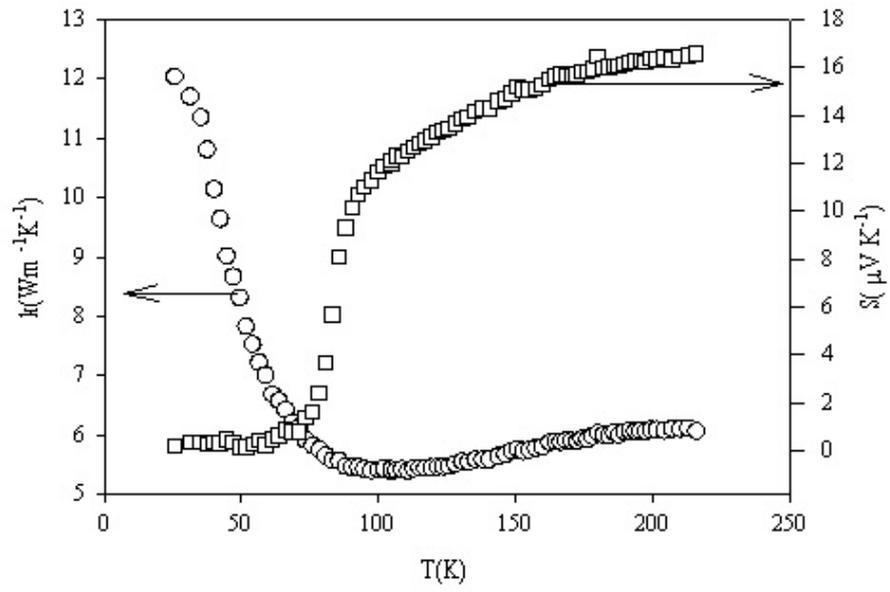

Fig.3

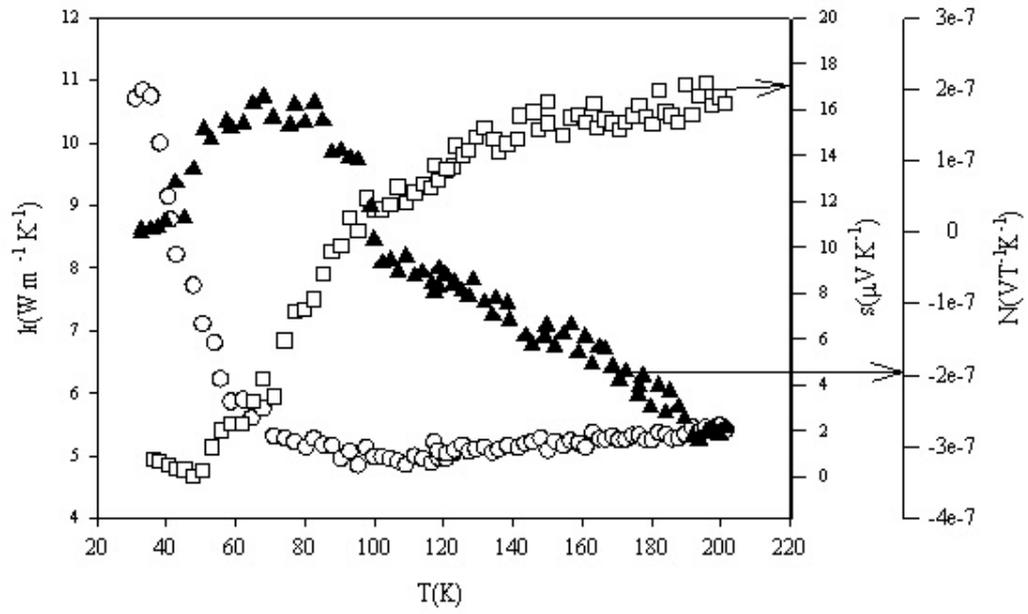

Fig.4